\documentclass[a4paper,twoside]{article}

\usepackage{epsfig}
\usepackage{subfigure}
\usepackage{graphicx}
\usepackage{caption}
\usepackage{svg}
\usepackage{calc}
\usepackage{amssymb}
\usepackage{amstext}
\usepackage{amsmath}
\usepackage{amsthm}
\usepackage{array}
\usepackage{multicol}
\usepackage{pslatex}
\usepackage{algorithmic}
\usepackage{titlesec}
\usepackage{balance}
\usepackage{apalike}
\usepackage{verbatim}
\usepackage{SCITEPRESS}     
\usepackage{algorithm2e}
\begin{document}

\title{OneDataShare:  \subtitle{A Vision for Cloud-hosted Data Transfer Scheduling and Optimization as a Service} }

\author{\authorname{Asif Imran, Md S Q Zulkar Nine, Kemal Guner, and Tevfik Kosar}
\affiliation{Department of Computer Science and Engineering, \\University at Buffalo, State University of New York, Buffalo, NY 14260}
\email{\{asifimra, mdsqzulk, kemalgne, tkosar\}@buffalo.edu}
}


\abstract{Fast, reliable, and efficient data transmission across wide-area networks is a predominant bottleneck for data-intensive cloud applications. This paper introduces \textit{OneDataShare}, which is designed to eliminate the issues plaguing effective cloud-based data transfers of varying file sizes and across incompatible transfer end-points. The vision of \textit{OneDataShare} is to achieve high-speed data communication, interoperability between multiple transfer protocols, and accurate estimation of delivery time for advance planning, thereby maximizing user-profit through improved and faster data analysis for business intelligence. The paper elaborates on the desirable features of \textit{OneDataShare} as a cloud-hosted data transfer scheduling and optimization service, and how it is aligned with the vision of harnessing the power of the cloud and distributed computing. Experimental evaluation and comparison with existing real-life file transfer services show that the transfer throughout achieved by \textit{OneDataShare} is  6.5 times greater.}

\onecolumn \maketitle \normalsize \vfill

\vspace{-4mm}
\section{INTRODUCTION}
\label{sec:introduction}
\vspace{-2mm}

Cloud services have gained tremendous popularity amongst the general public for its low cost, high availability, and elasticity. Millions of data files of varying sizes and formats are transferred from one storage to another every day using the cloud, irrespective of their locations. However, the benefits of cloud computing cannot be fully utilized due to the limitations in data transfer mechanisms, which have become a main bottleneck restraining full utilization of the cloud's strength. For example, transfer of a 1 TB dataset over to a cloud storage may take several weeks despite the high-speed networks available~\cite{garfienkel07}. For this reason, many IT companies and academic research labs prefer sending their data through a shipment service provider such as UPS or FedEx rather than using wide-area networks~\cite{cho11}. Some cloud storage providers (e.g., Amazon S3) even provide import/export services in which the users can ship multiple hard drives via FedEx to the storage provider, and the provider copies data to the cloud storage directly~\cite{stormbow}.
 
As data has become more abundant and data resources become more heterogeneous, accessing, sharing and disseminating these data sets become a bigger challenge. Using simple tools to remotely logon to computers and manually transfer data sets between sites is no longer feasible. Managed file transfer (MFT) services such as Globus Online~\cite{chard2017globus}, PhEDEx~\cite{egeland2010phedex}, Mover.IO~\cite{moverio}, B2SHARE~\cite{ardestani2015b2share}, RSSBUS~\cite{rssbus}, Stork~\cite{Royal_2011}, StorkCloud~\cite{ScienceCloud_2013}, and others~\cite{WORLDS_2004,IGI_2012} have allowed users to do more, but these services still rely on the users providing specific details to control this process, and they suffer from shortcomings including low transfer throughput, inflexibility, restricted protocol support, and poor scalability. 

 \begin{figure*}[t!]
     \centering
     \begin{minipage}[htp]{1.0\textwidth}
         {\subfigcapskip = 10pt 
         \subfigure[]{\includegraphics[angle=0,width=80mm, height=60mm]{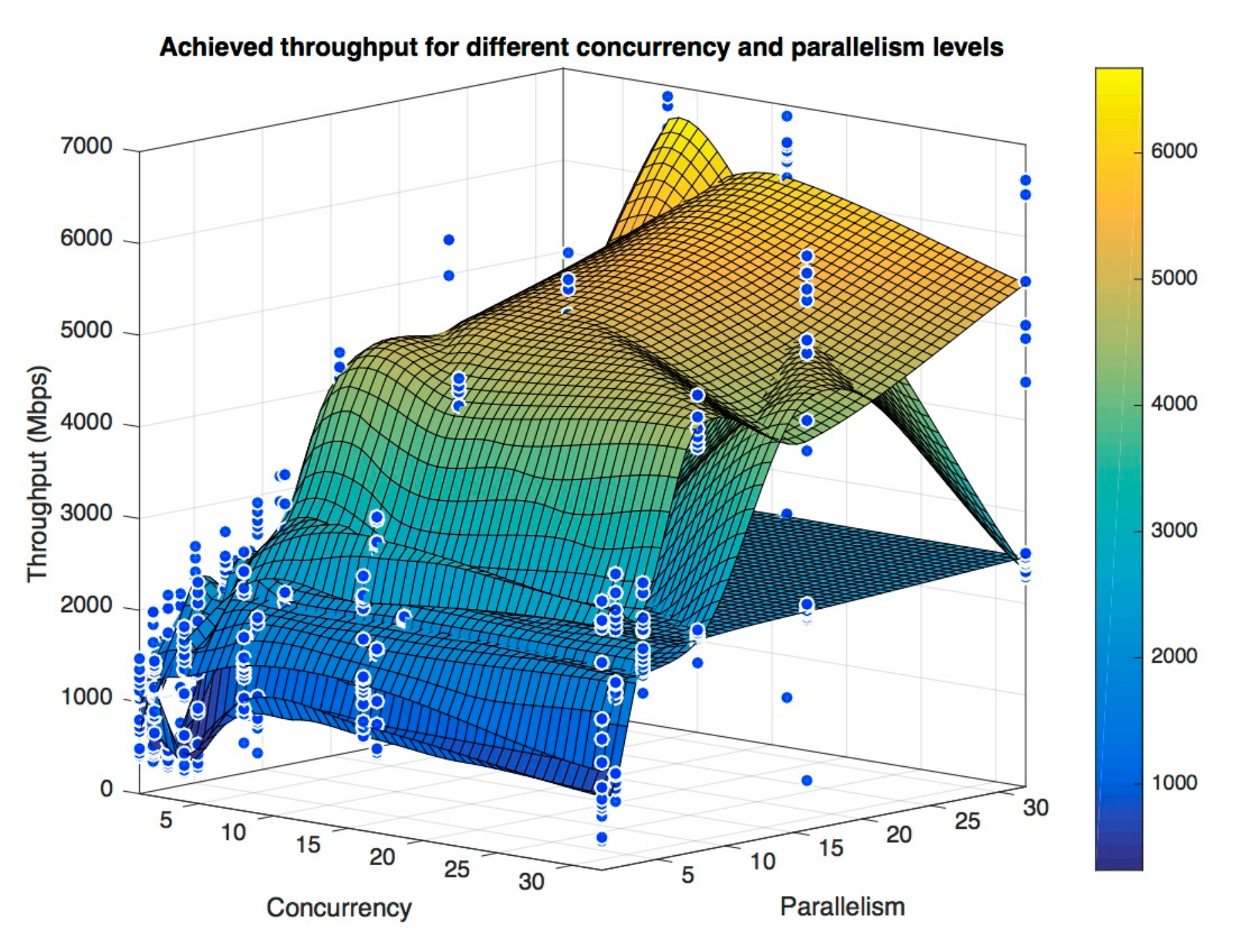}}}
         \subfigure[]{\includegraphics[angle=0,width=70mm, height=60mm]{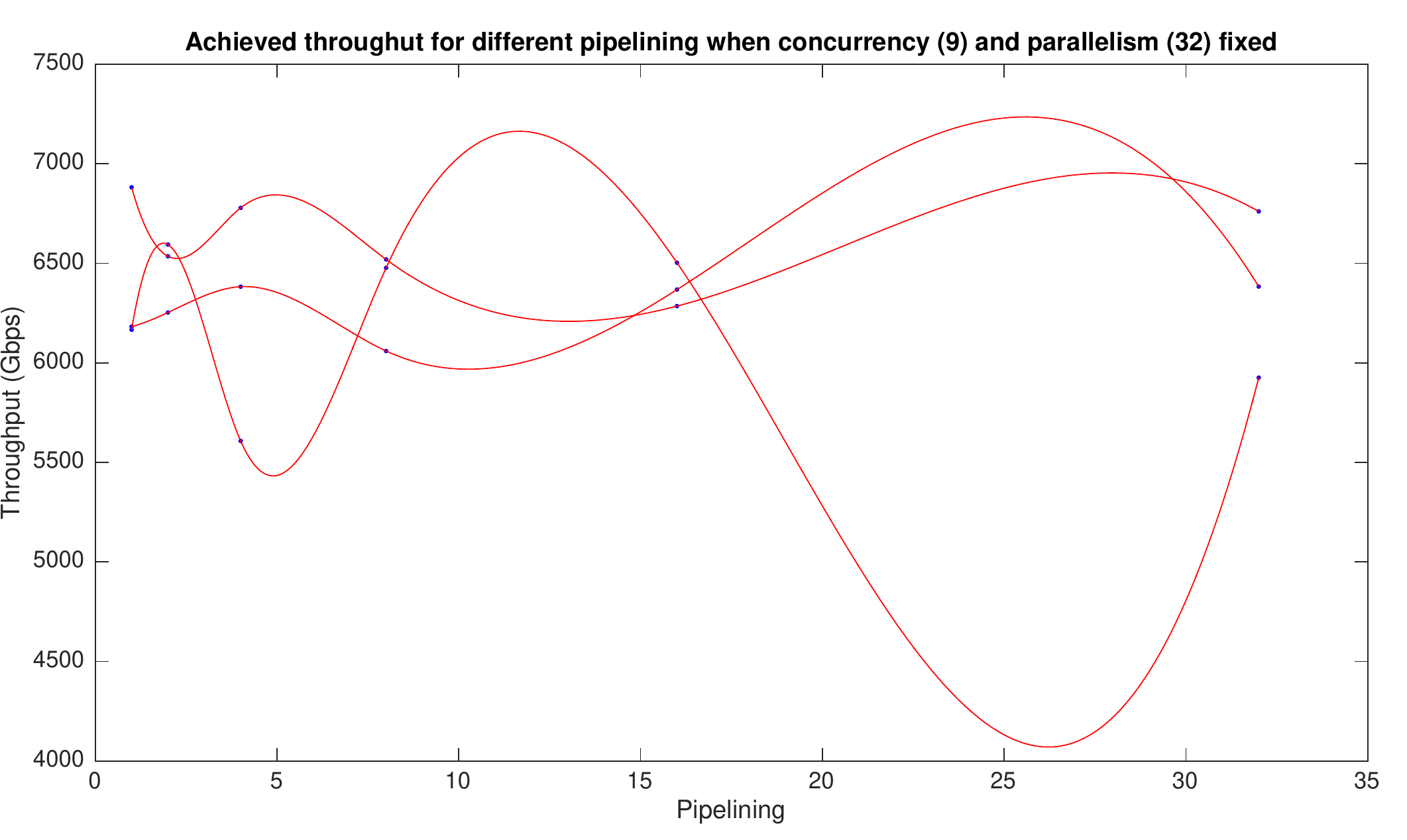}}
         \caption{ This figure shows the variation in achievable throughput in 10 Gbps XSEDE network between Stampede (TACC) and Gordon (SDSC). Cubic spline surface is constructed to interpolate throughput for the whole parameter space. (a) Shows the impact of concurrency and parallelism parameters on the throughput. The colors represent the throughput in Mbps, blue being the lowest and yellow being the highest throughput. The blue dots represent actual measured throughput values for the corresponding parameters. (b) The impact of pipelining on throughput is demonstrated.}
         \vspace{-2mm}
         \label{fig:thcubic}
     \end{minipage}
 \end{figure*}
 
Transferring large datasets especially with heterogeneous file sizes (i.e., small and large files mixed) causes inefficient utilization of the available network bandwidth. Small file transfers may cause the underlying transfer protocol not reaching the full network utilization due to short-duration transfers and connection start up/tear down overhead; and large file transfers may suffer from protocol inefficiency and end-system limitations. Application-level TCP tuning parameters such as 
pipelining~\cite{TCP_Pipeline,NDM_2011}, 
parallelism~\cite{DADC_2008,DADC_2009}, 
and concurrency~\cite{Thesis_2005,yildirim2012end} 
are very effective in removing these bottlenecks, especially when used together and in correct combinations. However, predicting the best combination of these parameters requires highly complicated modeling since incorrect combinations can either lead to overloading of the network, inefficient utilization of the resources, or unacceptable prediction overheads.

\begin{table*}[t]
\tiny
\begin{center}
 \begin{tabular}{||c | c | c || p{3cm} | c | c | p{5cm} | c ||} 
 \hline
 \multicolumn{3}{|c|}{\textbf{General Overview}} & \multicolumn{5}{|c|}{\textbf{Optimization information}} \\
 \hline
 \hline
 Name & Source & Cloud & Parameters & Fsize & Interface & Protocols & Metadata \\
 \hline\hline
 Globus & Open & Cloud & Parallelism, Pipelining, Concurrency & All & GUI & GridFTP, FTP, REST & Yes \\ 
 \hline
 PhEdEX & Open & Client & Compression & All & GUI & FTP, SCP, SFTP & Yes \\
 \hline
 Waarp & Open & Client & None & All  & Cmd line & FTP & No \\
 \hline
 DivConq & Open & Cloud & Compression & All & Cmd line & HTTP(S), SFTP & No \\
 \hline
 fileXhub & Open & Cloud & Distributed decoupling & Large & Cmd line & FTP, HTTP, JMS, SOAP, REST, SFTP & No \\
 \hline
 LDR & Open & Client & None & Medium & Cmd line & HTTP, FTP & No \\
 \hline
 Mover.io & Closed & Cloud & Compression, Parallelism & All & GUI & FTP, SFTP, REST, SMTP& Yes \\
 \hline
 IBM Aspera & Closed & Cloud & HDD striping, Compression, FASP & All & GUI & FASP & Yes \\
 \hline
 MoveIt & Closed & Client & CPU Freq, Parallelism, Pipelining & Medium & GUI & FTP, SFTP, HTTP & Yes \\
 \hline
 Gscape EFT & Closed & Client & None & All & GUI & FTP, HTTP, ASP & No \\
 \hline
 RSSBus & Closed & Cloud & HDD Stripe, TCP buffer & All & GUI & FTP, HTTP, & Yes \\
 \hline
\end{tabular}
\caption{Comparison among different managed file transfer and scheduling services.}
\label{t1}
\end{center}
\end{table*}

These tunable transfer parameters play a significant role in improving the achievable transfer throughput as shown in Figure \ref{fig:thcubic}. Here we can see, same transfer can achieve different throughputs for different parameter values. Figure~\ref{fig:thcubic}(a) shows the joint effect of concurrency and parallelism over a single transfer, where Figure~\ref{fig:thcubic}(b) shows the effect of pipelining. However, setting the optimal levels for these parameters is a challenging task and an open research problem. Poorly-tuned parameters can either cause underutilization of the available network bandwidth or overburden the network links and degrade the performance due to increased packet loss, end-system overhead, and other factors.

In this paper, we present the vision of \textit{OneDataShare} as a cloud-hosted data transfer scheduling and optimization service. OneDataShare has three major goals: {\em (i)} optimization of end-to-end data transfers and reduction of the time to deliver the data; {\em (ii)} interoperation across heterogeneous data resources (both streaming and at-rest) and on-the-fly inter-protocol translation; and {\em (iii)} predicting the data delivery time and decreasing the uncertainty in real-time decision-making processes.
OneDataShare's novel data transfer optimization, interoperability, and prediction services are being implemented completely at the application-level, not requiring any changes to the existing infrastructure nor to the low-level networking stack, although drastically increasing the end-to-end performance of data transfers and data-intensive applications depending on data movement. Experiments demonstrated that \textit{OneDataShare} achieved desirable throughput increase of 6.5 compared to existing data transfer services.

\vspace{-4mm}
\section{RELATED WORK}
\label{sec:relatedwork}
\vspace{-2mm}

The motivation for a platform independent cloud-to-cloud data transfer system has been specified in~\cite{aspera@2017}. Important gridlocks like local area storage bottleneck, WAN transport bottleneck, session restore limitation and slow throughput of data object transfers are limiting data transfer performance. Hence, Aspera~\cite{aspera2@2017} urged for development of a fundamental system that will cater to these issues. They identified some of the important characteristics which a cloud based data object transfer mechanism need to exhibit. Since the majority of cloud based storage are object storage, separation of file data and metadata together with replication across commodity storage are key to the success of effective data transfer~\cite{aspera@2017}. Also, there is a need to address the challenges faced when data objects are transferred across different continents via the cloud.  

Server to server data transfers in industry environments have been evaluated in \cite{carroll2017systems}. The proponents have analyzed what benchmarks should be followed when transferring data from one server to another at an industry level. They have evaluated and emphasized the importance of logging and time-stamping the transfer activity at every stage of the transfer for security and auditing purpose. Additionally, the authors provided a mechanism for identifying nearest server when the transfer needs to be speed effective. However, besides security and speed, how to support multi-protocol data transfers and design criteria for such a distributed data delivery system were not discussed. Also the overhead of time-stamping at every stage have been ignored.

Use of cloud platform to provide data transfer as a service to scientists with increased performance in terms of speed and security has been addressed in \cite{Allen_2012} which identifies Globus as a probable solution. Recent developments have focused on secured data transmission using Globus with transfer optimization using static parameters \cite{chard2017globus,liu2017explaining}. The authors showed results of data transfers using static parameters over the network using REST APIs, however Globus does not employ dynamic throughput optimization or interprotocol translation (at the time of the writing of this article). 

Scalability and data exchange capabilities of different MFT platforms have been discussed in \cite{singh2016managed}. Vertical and horizontal scaling systems have been suggested by the authors in order to address the issue of disk space overflow. \textit{Endpoint Scanning, Auto Recovery and Scheduler} modules have been proposed to address the need for constantly keeping the connectivity up and running to ensure glitch-free data transfer. Single point of scheduling server for multiple MFT communication, configuring one or more properties of servers and updating the entire system about it and using a centralized server for managing configurable job queue have been proposed. However, the issue of achieving effective software designs to incorporate patch updates at run time have been addressed to a limited extent.

Several MFT solutions present web based dashboards which allow users to specify the sending and receiving entities and also enable selection of the file which will be transferred over the network \cite{moverio}. However, many of these solutions are commercial and users need to purchase subscriptions in order to use these services for transferring the data. Also cross protocol data transfers together with optimization of the transfer itself are hardly addressed. Unfortunately, the commercial MFT services do not provide estimation of data arrival time, which is a critical information as it allows the recipient to prepare storage and data analytic tools. Also, automated protocol translation features have been incorporated to a limited extent. Finally, lack of robustness in these services is a common bottleneck to user-friendliness, which ultimately causes users to revert to traditional mechanisms of data transfers like FedEx. Table \ref{t1} provides a comparative analysis of the various Managed File Transfer (MFT) services currently in service from the perspective of functionality, usability and optimization capability.

\vspace{-4mm}
\section{ONEDATASHARE VISION}
\vspace{-2mm}

Efficient transfer of data is still a challenge despite the modern innovations in network infrastructure and availability of large bandwidth Internet \cite{yildirim2016application}. 
Being able to effectively use these high speed networks is becoming increasingly important for wide-area data exchange as well as for distributed and cloud computing applications which are data intensive.
For this purpose, we propose {\em OneDataShare} as a solution for universal data sharing with the following major goals:

\noindent {\bf (i) Reduce the time to delivery of the data.}
Large scale data easily generated in a few days may presently take weeks to transfer to the next stage of processing or to the long term storage sites, even assuming high speed interconnect and the availability of resources to store the data~\cite{NSF11}. Through {\em OneDataShare}'s application-level tuning and optimization of TCP-based data transfer protocols (i.e., GridFTP, SCP, HTTP), the users will be able to obtain throughput close to the theoretical speeds promised by the high-bandwidth networks, and the performance of data movement will not be a major bottleneck for data-intensive applications any more. {\em The time to the delivery of data will be greatly reduced, and the end-to-end performance of data-intensive applications relying on remote data will increase drastically.}

\noindent {\bf (ii) Provide interoperation across heterogeneous data resources.}
In order to meet the specific needs of the users (i.e., scientists, engineers, educators), numerous data storage systems with specialized transfer protocols have been designed, with new ones emerging all the time~\cite{shoshani2002storage}. Despite the familiar file system-like architecture that underlies most of these systems, the protocols used to exchange data with them are mutually incompatible and require specialized software to use. The difficulties in accessing heterogeneous data storage servers and incompatible data transfer protocols discourage researchers from drawing from more than a handful of resources in their research, and also prevent them from easily disseminating the data sets they produce.  {\em OneDataShare} will provide interoperation across heterogeneous data resources (both streaming and at-rest) and on-the-fly translation between different data transfer protocols. {\em Sharing data between traditionally non-compatible data sources will become very easy and convenient for the scientists and other end users.}

\noindent {\bf (iii) Decrease the uncertainty in real-time decision-making processes.}
The timely completion of some compute and analysis tasks may be crucial for especially mission-critical and real-time decision-making processes. If these compute and analysis tasks depend on the delivery of certain data before they can be processed and completed, then not only the timely delivery of the data but also the predictive ability for estimating the time of delivery becomes very important~\cite{DOE13}. This would allow the researchers/users to do better planning, and deal with the uncertainties associated with the delivery of data in real-time decision making process. {\em OneDataShare}'s data throughput and delivery time prediction service {\em will eliminate possible long delays in completion of a transfer operation and increase utilization of end-system and network resources by giving an opportunity to provision these resources in advance with great accuracy.} Also, this will enable the data schedulers to make better and more precise scheduling decisions by focusing on a specific time frame with a number of requests to be organized and scheduled for the best end-to-end performance.

While realizing these goals, {\em OneDataShare} will make the following contributions to the distributed and cloud computing community:
{\em  (i)} implementation of novel and proven techniques (online optimization based on real-time probing, off-line optimization based on historical data analysis, and combined optimization based on historical analysis and real-time tuning) for application-level tuning and optimization of the data transfer protocol parameters to achieve best possible end-to-end data transfer throughput;
{\em (ii)} development of a universal interface specification for heterogeneous data storage endpoints and a framework for on-the-fly data transfer protocol translation to provide interoperability between otherwise incompatible storage resources;
{\em (iii)} instrumentation of end-to-end data transfer time prediction capability, and feeding of it into real-time scheduling and decision making process for advanced provisioning, high-level planning, and co-scheduling of resources; and
{\bf (iv)} deployment of these capabilities as part of stand-alone {\em OneDataShare} cloud-hosted service to the end users with multiple flexible interfaces.


\vspace{-4mm}
\section{ONEDATASHARE DESIGN}
\vspace{-2mm}

\textit{OneDataShare} will add significant value to the field of cloud based data transfers due to its adoption of elasticity and sustainability. However, significant features have been specified in \textit{OneDataShare} architecture to ensure that the solution adheres to its promised performance. Incorporation of these features is not only performance critical, but also vital to the acceptance of \textit{OneDataShare} by the general users. 

\begin{figure}[t]
  \centering
   \includegraphics[width=75mm,height=90mm]{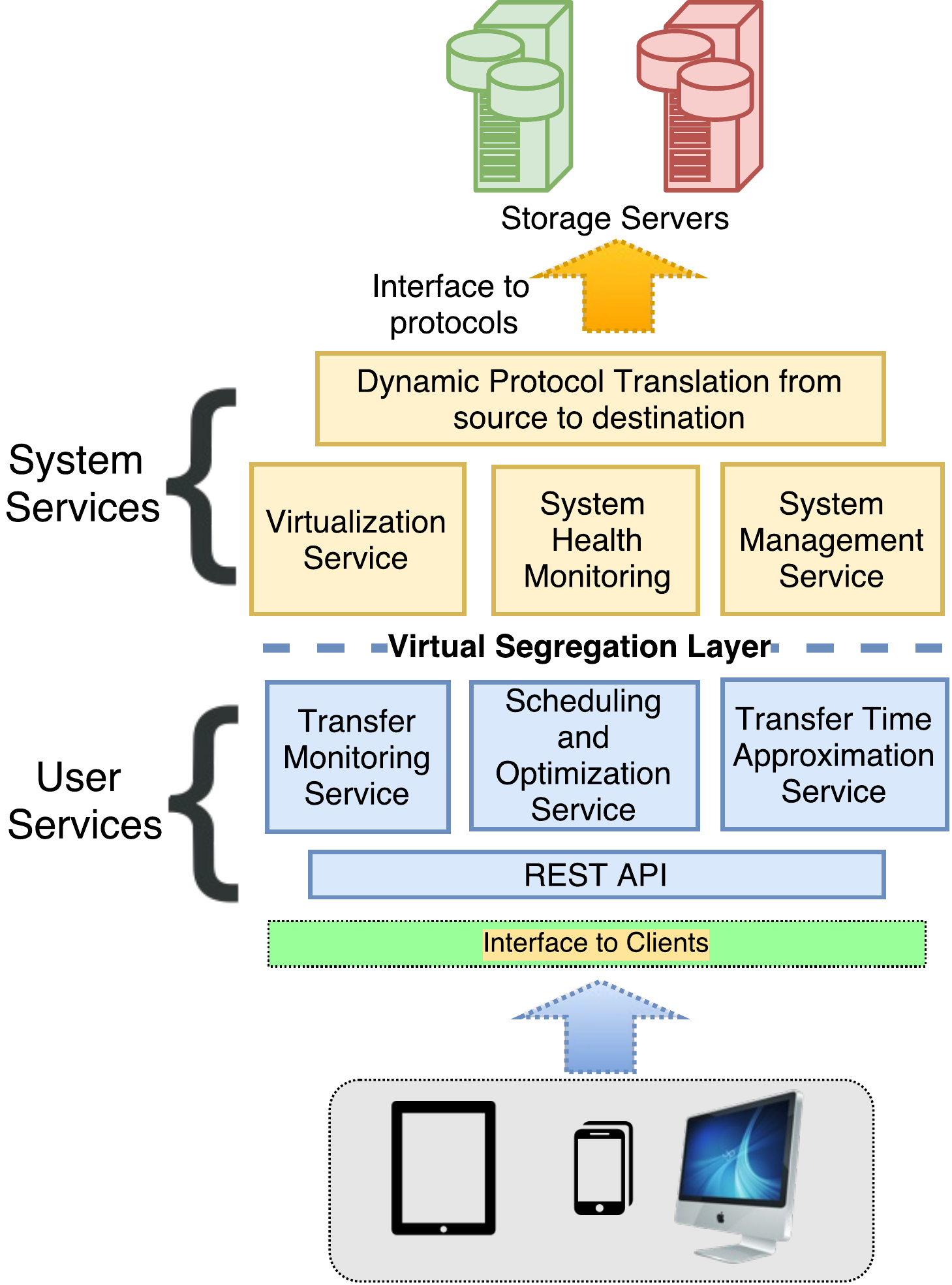}
  \caption{OneDataShare high-level overview.}
  \label{toplevel}
\end{figure}

This section identifies some of the most important key design features of \textit{OneDataShare}. Figure \ref{toplevel} provides a visual representation of \textit{OneDataShare}, showing the interim protocols and mechanisms in place and how it achieves improved data sharing and transfer optimization while providing a thin client interface to the users. \textit{OneDataShare} is designed to support a variety of interfaces for the users which include web interface, smartphone and tablet apps, as well as command-line and file system interfaces. When a user requests for a transfer service to \textit{OneDataShare}, the request is submitted to the engine of the service which contains RESTful service with a myriad collection of schedulers, protocol translators, provenance managers and cloud manager. This complex and dynamic collection of modules appears as a black box to the general users of \textit{OneDataShare}. Also, the entire protocol translation, optimization and scheduling of arrival time of the data files is maintained in the cloud, thus ensuring reliability and high availability of service to the users. It is a ubiquitous service as users will only have to pay for the resource requirements of their specific requests, thereby saving unnecessary costs. The status of data transfers and overall health of the internal components are monitored by the \textit{System Monitor} module.

Based on the dynamic nature of \textit{OneDataShare}, it must cater to the demands of the users regarding transfer of data which involves high speed of data delivery, interoperability, and improved estimation of delivery time
which are discussed below. 

\subsection{High Speed Data Delivery} 
\vspace{-2mm}

The most desirable feature of \textit{OneDataShare} is high speed delivery of the data through application-level protocol tuning using the underlying network architecture. 
Prior work on application level tuning of transfer parameters mostly proposed static or non-scalable solutions to the problem with some predefined values for some generic cases~\cite{Allen_2012,R_Hacker02,R_Crowcroft98,R_Dinda05}. The main problem with such solutions is that they do not consider the dynamic nature of the network links and the background traffic in the intermediate nodes. 
In our previous work, we have developed three highly-accurate predictive models ~\cite{R_Yin11,R_Yildirim11,DISCS12} which would require as few as three real-time sampling points to provide very accurate predictions for the optimal parallel stream number. These models have proved to have higher accuracy compared to existing similar models \cite{R_Hacker02,R_Dinda05} which lack in predicting the parallel stream number that gives the peak throughput. We have analyzed the combined effect of these transfer parameters on end-to-end data transfer throughput, and developed several predictive (offline) and dynamic (online) algorithms to choose the best parameter combination to minimize the delivery time of the data~\cite{yildirim2016application,Europar2013,jkim15,yildirim2012end,Arslan2016}.


Heuristic approaches are solely based on domain knowledge, another approach could be the acquisition of such knowledge directly from the historical data transfer logs. Heuristic approaches might over-generalize the prediction for some systems, where historical analysis based approaches actually mine useful knowledge from the user data transfer pattern, end nodes and their specific parameter values of connecting link. Such approach can provide more personalized optimization for the users and the corresponding systems. We have collected production level data transfer logs from XSEDE, a well-known infrastructure for high performance scientific computing. Those transfer logs contain information about end systems, dataset, network links, and the protocol along with parameter settings. Proper data mining techniques on such a rich collection of historical logs should reveal interesting user transfer pattern, intensity of network traffic (e.g. peak or off-peak hours), and end system specific suitable parameter settings. Even if the data transfer logs are very detailed, it is not possible to collect information for a transfer with many different parameter settings. Only a partial view of the whole parameter space can be extracted. As the partial view of parameter space might not contain the optimal settings, we can use regression or interpolation techniques to predict more detailed view of the parameter space. 

Another important aspect of data transfer optimization is the dynamic load change in the network link. Even if the historical analysis provide more fine-grained parameter tuning than heuristics based approaches, still it might prove sub-optimal for a network load that is different from network load present in the historical logs. Instead of transferring the whole dataset using parameters from historical analysis, some portion of the dataset could be transferred to assess the network load intensity and tuned the parameters accordingly should improve the overall throughput. However, deciding the appropriate sample transfer size and numbers is a critical research issue.   Sometimes network load changes during the large data transfer. Therefore, dynamic tuning of the parameters on the fly will be a very useful feature in \textit{OneDataShare}.

We introduced historical analysis based approach in ANN+OT~\cite{Nine_2015} that uses machine learning techniques to learn optimal parameters from the historical logs. In this model we have used machine learning based techniques to get optimal parameters for different types of network. We have used Artificial Neural Networks and Support Vector Machines, two well known supervised learning techniques, to learn optimal parameters from the transfer logs. This approach provided considerable improvement over the existing techniques. Even though historical analysis can provide more accurate parameters than heuristic approaches, the dynamic nature of the network might prove those parameters sub-optimal. Therefore, current network condition is also an important factor to set tuning parameters. We have made some progress in real-time sampling in ANN+OT. It performs a series of real-time sampling to assess the current network condition and update the protocol parameters accordingly.

\begin{figure}[t]
  \centering
  \includegraphics[width=7cm,height=4cm]{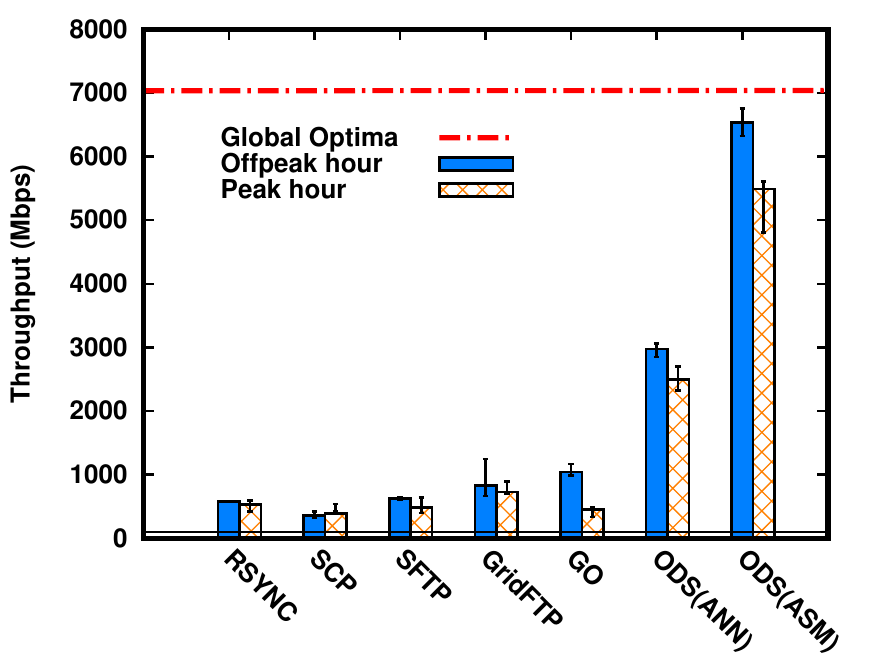}
  \caption{Performance comparison between different data transfer services for both peak and off-peak hours.}
  \label{fig:model_comparison}
\end{figure}

In our most recent work~\cite{Nine_2017}, we introduced a two phase model aimed to reduce the performance degradation due to sampling overhead. 
It uses a robust mathematical model based offline analysis on the historical logs to interpolate the throughput surface for the parameter space. It stores the most interesting regions of the surface and local maxima points for different network conditions. During online phase, instead of performing sample transfers blindly, it adapts the parameters using the guidelines from the offline analysis to achieve faster convergence.  

In an effort to compare OneDataShare performance to other state-of-the-art solutions, we run an experiment in which we transferred data using the production level XSEDE nodes on Stamepede2 and Comet. We transferred the data using different file transfer services, such as - scp, rsync, sftp, GridFTP, Globus Online (GO) and compared those results with our OneDataShare(ODS) service with two optimization model - ANN+OT and ASM as shown in Figure \ref{fig:model_comparison} where we refer them as ODS(ANN) and ODS(ASM) respectively. Here we can see most of the current data transfer services achieves fraction of the achievable throughput due to their sub-optimal protocol parameter choices. scp and rsync are very popular file transfer tools, however, performance is quite marginal. SFTP protocol itself has many performance issues for long RTT high speed networks. GridFTP overcomes many issues of FTP based protocols, such as data channel reuse, pipelining, parallel streams, concurrent file transfers~\cite{R_Allcock05,NDM_2012}. We can see GridFTP and Globus Online performs better than scp, rsync or sftp. However, performance can be increased by tuning the protocol parameters, like - pipelining, parallelism, concurrency, tcp-buffer size etc. In our proposed OneDataShare we used two highly tuned protocol optimization techniques- ANN+OT and ASM. ODS(ANN) shows $3 \times$ performance increase compared to GO, where ASM achieves almost $6.5 \times$ throughput increase. 

\subsection{Interoperability} 
\vspace{-2mm}



\begin{figure}[t]
  \centering
  \includegraphics[width=7cm,height=4cm]{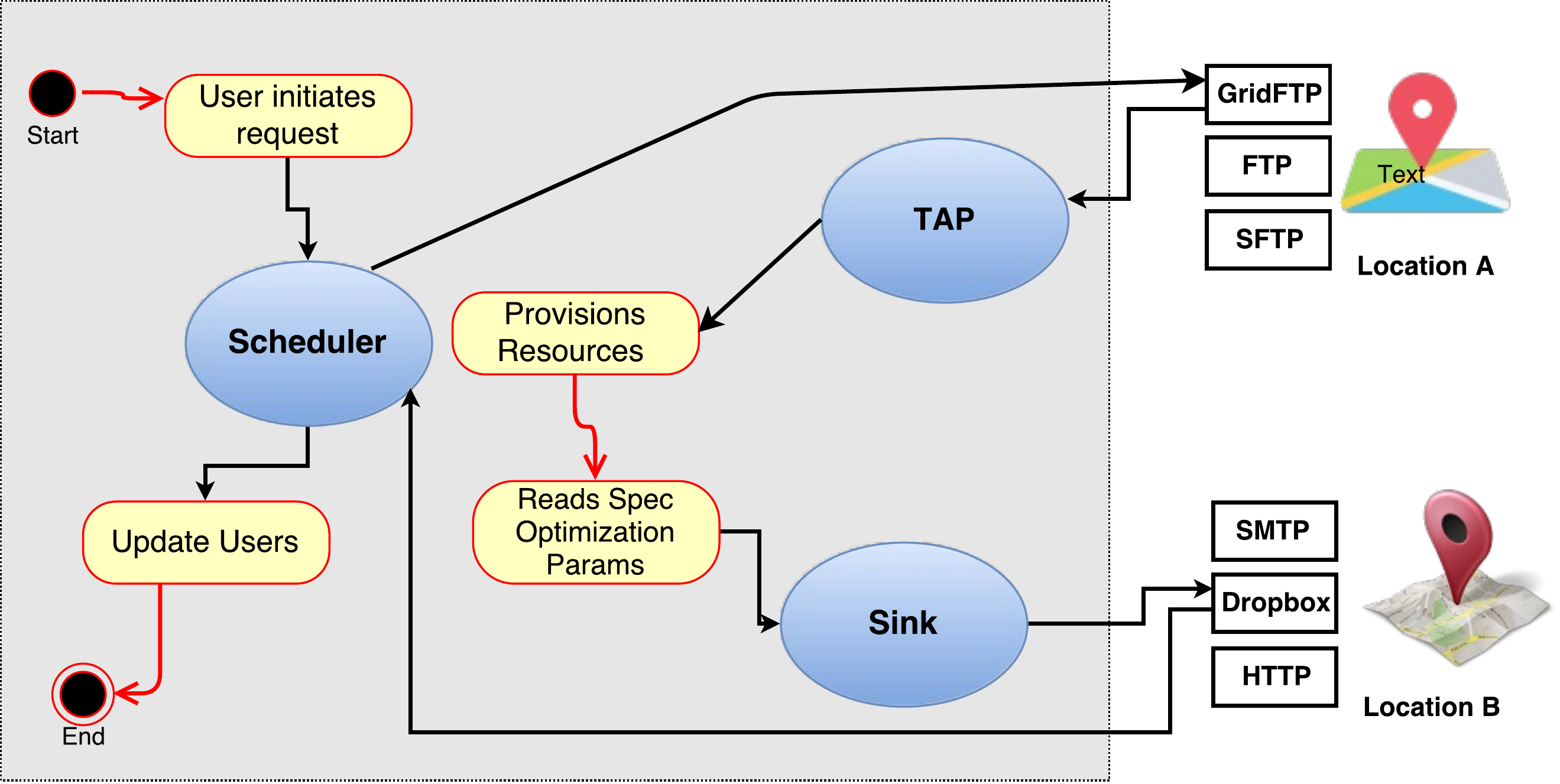}
  \caption{OneDataShare's interoperability components.}
  \label{onedshare}
\end{figure}

\textit{OneDataShare} ensures that data sent using \textit{Protocol X} can be delivered at the recipient in a different protocol (i.e. Protocol Y) without much effort from the user end. This protocol translation mechanism is a black box to the user and implements a fast, effective and timely solution to on-the-fly protocol transformation.
In Figure \ref{onedshare}, it is seen that a user from Location A wants to transfer a file to a system at Location B using GridFTP protocol. However, user at location B will accept the data file in its Dropbox. Additionally, Figure \ref{onedshare} exhibits \textit{OneDataShare}'s mechanism to achieve this protocol translation on-the-fly. 

\textit{OneDataShare} must support a wide array of protocols to cater to the ever increasing demand of cross protocol transfer of data in the cloud. Since the cloud can host data of all formats and sizes, \textit{OneDataShare} needs to possess the provision of multiple protocols. Also, the software must be flexible and sustainable enough to incorporate new protocols without service stoppage since new protocols are introduced any time as an outcome of continuous research in this area. 

Placing the burden of protocol translation on the user will provide extra burden on them. The users may not be adept to protocol translation mechanisms as they are not used to technological perspectives of converting data from one protocol to the next. It is very common that large volumes of data are transferred by the users with backgrounds of non computing knowledge, hence, carrying out protocol translation will impose burden on them and discourage them to use the network for transferring data. 

Here, the state of the art \textit{Tap} and \textit{Sink} mechanisms are employed for protocol translation as shown in Figure \ref{onedshare} which is described later in this paper. 
When the process is passed through the protocol translation gateway and the server is triggered, instantaneously the server reads the incoming file properties and contents, analyzes the incoming protocols used in the tap and writes the data in memory. After that the data bytes are re-commissioned  into a desirable format of the protocol accepted at the recipient which is Dropbox as highlighted in Figure \ref{onedshare}. Finally, it is seen that the file is securely delivered. 


In this framework, the readable resources implement the {\em Tap} operation to acquire a data {\em tap} which will “emit” data into a data {\em sink}; and the write-able resources implement {\em Sink} operation to acquire a data {\em sink} which will “drain” data from a data {\em tap}.

The life cycle of \textit{OneDataShare} inter-protocol translation begins with the user triggering a transfer to the transfer manager known as \textit{tap}. The \textit{tap} is one of the core modules which is responsible for accessing the file's source and prepare for protocol translation for successful transfer of the file to its destination.

The \textit{tap} is the frontier interface, responsible for abstracting the back-end components of \textit{OneDataShare} which may pose significant complexity if exposed to general users. The \textit{tap} provisions a unified single point opening wedge to the system. The UI is a dashboard which presents all the functionalities which can be harnessed by the users. It also allows users to initiate transfer requests.

The dashboard is also responsible for transmitting the selected functionalities and optimization to the back-end engine. Afterwards, provisioning of the computing and network resources are done. The system obtains authorized file access permissions and loads all child directories from which the transfer files may be selected. Next, the \textit{tap} receives this input and ultimately enables the deployment of the process which places the data files into he provisioned resources, attaches protocol related information which primarily constitutes of the names of source and destination protocols and transmits these journals to \textit{sink}.

The \textit{sink} is the delivery manager of data which triggers the recipients of the incoming requests and prepares them for receiving this data. It is a self-protocol-translating system which takes input from the \textit{tap} and formats the data into forms which is suitable to be accepted at the recipient. It envisages the incoming data and loads it in memory, typically associating them to the resulting protocol, embedding them and mitigating the network transfer requirements based on the optimization parameters as specified by users. The \textit{sink} is empowered with multitude of child processes which cater to the requirements of multiple block data transfers per unit time.

More specifically, \textit{sink} is fully fitted with the required processes of computing, memory and network resources which are used during protocol translations and initiating data transmissions. These resources are provisioned dynamically using cloud computing and the users are billed ubiquitously based on their usage. The elastic capabilities of the cloud is harnessed here as when it is read from journal files that resource demand will increase due to sudden rise in the number of transfer requests, additional resources can be added dynamically and the system is automatically scaled up as per the demand.

The advantage of this approach is that existing applications and storage systems do not need to be changed to be part of this framework. This is particularly important as many tools and libraries used in scientific
applications cannot be changed easily to accommodate new storage systems.
Introducing this capability will require development of an abstraction model which appropriately encapsulates storage system server functionality; which will in turn be implemented in the existing transfer modules.
This solution presents a number of strengths over other solutions: {\em i)} neither servers nor clients need to install custom software to take advantage of the protocol translation {\em OneDataShare} offers;  {\em ii)} support for additional protocols can be added on-the-fly
without having to make changes to any other system. 
Through this framework, {\em OneDataShare} will provide interoperability and on-the-fly protocol translation between a wide-range of data transfer protocols and storage systems, including but not limited to FTP, GridFTP, HTTP, SCP, SFTP, Rsync, BBFTP, UFTP, UDT, iRODS, SMTP, Dropbox, and Google Drive.

\subsection{Transfer Time Estimation} 
\vspace{-2mm}

It is seen that in the modern world that both scientific research organizations and industries are struggling to estimate the time of arrival of data after it has been transferred. This information is critical as the arrival of data will require provisioning storage for the data itself at the recipient's end. Additionally, the algorithms and software components which will be used to analyze these data need to be prepared before it arrives, and compute resources need to be provisioned accordingly. 
The timely completion of compute and analysis tasks are especially crucial for mission-critical and real-time decision-making processes. If these tasks depend on the delivery of certain data before they can be processed, then not only the timely delivery of the data but also the predictive ability for estimating the time of delivery becomes very important. This allows the researchers/users to do better planning, and to deal with the uncertainties associated with the delivery of data in real-time decision making process.

If storage and compute resources are provisioned long time before the arrival of data then the client needs to pay for the time during which these resources were provisioned but left idle because the data did not arrive. Hence, estimation of arrival time accurately will provide benefits to the users in a multitude of ways which includes time and cost benefits. \textit{OneDataShare} will use dynamic prediction algorithms to estimate arrival time of data to a significant degree of accuracy, thereby allowing the users to provision both storage, computing, software and human resources on time for using the data. It principally ensures that the user is well aware and they can plan ahead to better utilize the resources and analyze the data. 

Using the prediction models which we have developed in our previous work, {\em OneDataShare} will provide an ``end-to-end data transfer throughput and delivery time estimation service"  for external data scheduling and management tools. Our prior work on predictive models showed that we can estimate the real-time achievable throughput with as low as 5\% error rate on average~\cite{jkim15,yildirim2012end,yildirim2013modeling}. This service will include information regarding available end-to-end network throughput for the user, the total time it will take to transfer a particular dataset, network and end-system parameters that need to be used in order to achieve highest end-to-end throughput. All this information can help users to evaluate the network and data resource conditions before and during each data transfer, and not only reduce the time to delivery of the data but also allow predicting the time when the data may arrive at the recipient so that users of the data at the scientific and industry levels can deal with uncertainty in real-time decision making processes.

\vspace{-4mm}
\section{CONCLUSIONS}
\label{sec:conclusion}
\vspace{-2mm}

The vision of \textit{OneDataShare} has been presented here which aims to remove the burden related to data transmission from the shoulders of users. It en-visions the goal of providing a simple yet user friendly platform to share data irrespective of size, type, and difference in protocols between sender and receiver. The desirable features of such a system are discussed here and how \textit{OneDatashare} integrates these features are explained through design analysis which is provided in this paper. Real life experiments have been conducted to exhaustively compare performance of \textit{OneDataShare} with other services and it is seen that the data transfer throughput of \textit{OneDataShare} is 6.5 times greater than other commercial systems currently in place. Building \textit{OneDataShare} into a completely reliable and sustainable application beneficial to all is an area of future research. 

\vspace{-4mm}
\section*{ACKNOWLEDGEMENTS}
\vspace{-2mm}

This project is in part sponsored by the National Science Foundation (NSF) under award number OAC-1724898.
We also would like to thank Brandon Ross, Vandit S. Aruldas, Mythri Jonnavittula, Ahmad M. Temoor, Dhruva K. Srinivasa, Jimmy Huang, Sourav Puri, Siddharth Ravikumar, and Amardeep Virdy for their contributions in the implementation of this project.

\bibliographystyle{apalike}
{\small
\bibliography{main}}

\vfill
\end{document}